\documentclass[pra,twocolumn,showpacs]{revtex4}
\usepackage{amsmath,amssymb,graphicx}
\newcommand{\be}{\begin{equation}}
\newcommand{\ee}{\end{equation}}
\newcommand{\vars}{(z_i,\bar{z}_f,T)}

\newcommand{\bz}{\bar{z}}
\newcommand{\w}{\omega}
\newcommand{\ket}{|z\rangle}
\newcommand{\bra}{\langle z|}

\begin{document}
\title{Semiclassical propagation of spin coherent states}
\author{Marcel Novaes}
\affiliation{Instituto de F\'{i}sica ``Gleb Wataghin",
Universidade Estadual de Campinas, 13083-970 Campinas, S\~ao
Paulo, Brazil}

\begin{abstract}
The semiclassical propagation of spin coherent states is
considered in complex phase space. For two time-independent
systems we find the appropriate classical trajectories and show
that their combined contributions are able to describe quantum
interference with great accuracy. Not only the modulus but also
the phase of the quantum propagator, both dynamical and geometric
terms combined, are accurately reproduced.
\end{abstract}


\maketitle

\section{Introduction}
The coherent state path integral for spin systems and its
semiclassical approximation have appeared simultaneously
\cite{prd19jrk1979,jmp21hk1980}, and since then they have been
intensively studied. Solari \cite{jmp28hgs1987}, and independently
Kochetov \cite{kochetov,koche2}, have shown the existence of an
initially unexpected term that is sometimes called the
Solari-Kochetov ``extra phase", even though it is usually not a
phase. Some important applications of spin path integral were in
the study of spin tunnelling in the semiclassical limit
\cite{prl60emc1988,prl69dl1992,prb45ag1992}, even though initially
some considered the method to be inaccurate \cite{inac}. Stone
{\it et al} \cite{jmp41ms2000} and also Vieira and Sacramento
\cite{vieira} have derived the spin coherent state semiclassical
propagator in detail, paying particular attention to the
Solari-Kochetov correction. This correction is related to the
difference between the average value of the Hamiltonian in
coherent states and its Weyl symbol \cite{jmp45mp2004}, and has a
counterpart in the canonical case \cite{koche2,jpa34mb2001}, which
has a flat phase space. The semiclassical description of tunneling
was reconsidered lately, using the instanton method
\cite{instanton,jumps}.

Recently, a semiclassical quantization condition for one
dimensional spin systems has been derived \cite{prl92ag2004},
including the first quantum corrections, in the spirit of the
Bohr-Sommerfeld formalism. The same quantization condition was
reobtained in \cite{pra71mn2005}, where the authors also presented
a semiclassical expression for the Husimi functions of stationary
states. Semiclassical theories for spin-orbit coupling in
connection with trace formulas have appeared \cite{spinorbit1} and
recently received renewed attention \cite{spinorbit2}.

It is interesting to note that the coherent state spin path
integral is a natural setting for the investigation of geometric
phases \cite{geom,non}, a topic that has not only a fundamental
importance in the mathematical structure of quantum theory
\cite{mat} but also as an ingredient in the implementation of some
quantum information protocols \cite{qip}. The semiclassical
approximation to the spin propagator $K\vars=\langle
z_f|e^{-iHT/\hbar}|z_i\rangle$ may sometimes lead to an expression
to the geometric phase in terms only of classical quantities.

In this work we present a concrete application of the
semiclassical spin propagator. In general the calculation involves
a classical trajectory, $(z(t),\bz(t))$, that starts at $z=z_i$
and ends at $\bz=z_f^\ast$ after a time $T$. Under these too
stringent conditions, the only way to find a classical trajectory
is by allowing the variable $\bz(t)$ to be different from the
complex conjugate of $z(t)$ (which is denoted by $z^\ast(t)$). We
must therefore find a trajectory in $\mathbb{C}^2$ that satisfies
the boundary conditions $z(0)=z_i$ and $\bz(T)=z_f^\ast$. This is
known as the root-search problem. Similar calculations have
already appeared for the canonical coherent states
\cite{canonical1}, even in chaotic cases \cite{pre69adr2004}, but
so far no numerical example for a spin system has been presented.

The article is divided as follows. In the next section the
semiclassical theory of the spin coherent states propagator is
briefly reviewed. The geometric phase and the tangent matrix
associated with classical trajectories are also presented. In
section III we study the simple system $\hat{H}=\hbar^2\nu J^2_z$,
which already has nontrivial properties. In section IV the less
symmetrical Hamiltonian $\hat{H}=\hbar^2\nu[J^2_z+\pi J^2_x]$ is
considered, and in section V we present the conclusions.

\section{The semiclassical propagator}

Let $|z\rangle$ denote non-normalized spin coherent states,
defined by
\be\label{states}|z\rangle=\exp\{zJ_+\}|-j\rangle=\sum_{m=-j}^j\begin{pmatrix}
  2j \\
  m
\end{pmatrix}^{\frac{1}{2}}z^{j+m}|m\rangle,\ee where the states $|m\rangle$ are the usual spin basis,
and let \be \label{prop}K\vars=\langle
z_f|e^{-i\hat{H}T/\hbar}|z_i\rangle,\ee be the quantum propagator,
where $\hat{H}$ is the spin Hamiltonian and $T$ is the time. It
has been shown that, in the semiclassical limit $j\to\infty$,
$\hbar=1/j\to 0$, this can be approximated by
\cite{jmp41ms2000,vieira} \be\label{propag} K_{\rm
sc}\vars=\sum_{\rm c.t.} \left(\frac{i}{\hbar}\frac{e^{iB/\hbar
j}}{2j}\frac{\partial^2 S}{\partial z_i\partial
\bz_f}\right)^{\frac{1}{2}}\exp
\left\{\frac{i}{\hbar}\Phi\right\},\ee where the sum is over
different classical trajectories. The exponent is the classical
action $S$ plus an extra term, the Solari-Kochetov (SK)
correction: \be\label{phi} \Phi= S+\mathcal{I}_{\rm SK}
=S+\int_0^TA(t)dt.\ee It is well known that this semiclassical
approximation is exact if the Hamiltonian belongs to the $su(2)$
algebra \cite{kochetov,koche2,vieira,jmp41ms2000}.

The function $A$ is given by \be A=\frac{\partial}{\partial
\bz}\frac{1}{4g(z,\bz)}\frac{\partial H}{\partial
z}+\frac{\partial}{\partial z}\frac{1}{4g(z,\bz)}\frac{\partial
H}{\partial \bz},\ee where $H(z,\bz)=\bra \hat{H}\ket/\bra
z\rangle$ plays the role of the classical Hamiltonian and the
metric factor is \be g(z,\bz)=\frac{\partial ^2}{\partial z
\partial \bz}\ln\bra z\rangle=\frac{2j}{(1+z\bz)^2}.\ee
The classical action is given by \be\label{action} S=\int_0^T
\left[i\hbar j\frac{\bz \dot{z}-\dot{\bz}z}{1+\bz
z}-H(z,\bz)\right]dt+\mathcal{B}, \ee where $\mathcal{B}=-i\hbar j
\ln[(1+\bz_fz(T))(1+\bz(0)z_i)]$ is a boundary term, that takes
into account the fact that in general $\bz$ is not the complex
conjugate of $z$.

The integrals in (\ref{phi}) and (\ref{action}) are to be
calculated along classical trajectories satisfying the Hamilton
equations of motion \be
\dot{\bz}=\frac{i}{\hbar}\frac{1}{g(z,\bz)}\frac{\partial
H}{\partial z}, \quad
\dot{z}=-\frac{i}{\hbar}\frac{1}{g(z,\bz)}\frac{\partial
H}{\partial \bz},\ee with boundary conditions \be z(0)=z_i, \quad
\bz(T)=z_f^{\ast}.\ee Notice that, in general, \be z(T)\neq z_f,
\quad \bz(0)\neq z_i^{\ast}.\ee Since $\bz(t)\neq z^\ast(t)$ the
Hamiltonian $H(z,\bz)$, the action and the SK correction will in
general all be complex numbers. Therefore the term
$\exp\{i\hbar^{-1}\Phi\vars\}$ is usually not just a phase.

If the propagator is written as a modulus times a phase,
$K\vars=|K|e^{i\varphi}$, it is well known that $\varphi$ contains
not just the dynamical term, $-i\int_0^THdt$, but also the
geometric (or Berry) phase $\varphi_g$, which depends only on the
geometry of the path traced by the state in the Hilbert space
\cite{geom,mat}. This was shown by Berry in the context of
Hamiltonians depending on slowly varying cyclic parameters, and
later generalized in many different ways, in particular to
parameter independent, non-adiabatic and non-cyclic evolutions
such as the one considered here \cite{non}. From equation
(\ref{propag}) we see that in the semiclassical limit the total
phase will result from the interference of many classical
trajectories, and that each one of them has an individual phase
which is the sum of a dynamical term plus a geometric term given
by \be\label{geo} \varphi_p+\frac{1}{\hbar}{\rm Re}
\left\{\mathcal{B}+\int_0^T \left[i\hbar j\frac{\bz
\dot{z}-\dot{\bz}z}{1+\bz z}+A(t)\right]dt\right\},\ee where
$\varphi_p$ is a phase coming from the prefactor.

If we set $z_f=z_i$ in the simplest case $\hat{H}=\nu \hbar J_z$,
we have $z(t)=e^{-i\nu t}z_i$ and $\bz(t)=e^{i\nu (t-T)}z_i^\ast$,
and there is only one classical trajectory. The stability matrix
element $M_{\bz\bz}$ (see below) in this case is simply $e^{i\nu
T}$. For the very particular time $T=2\pi/\nu$ it happens that
$\bz(t)=z^\ast(t)$ and thus $\mathcal{B}$ is purely imaginary. If
we use stereographic coordinates $z=e^{i\phi}\tan(\theta/2)$ then
it is easy to see that, since $A$ and $\varphi_p$ both vanish, the
geometric phase reduces to the well known result $ \varphi=2\pi
j(1-\cos\theta)$. Generically, i.e. if $\hat{H}=\nu J_z$ but $T$
is not a multiple of the period, or for more general Hamiltonians,
the fact that $\bz(t)\neq z^\ast(t)$ will introduce additional
contributions coming from $A$, $\varphi_p$ and $\mathcal{B}$, even
for cyclic evolutions. Besides, more than one classical trajectory
will be necessary, as we will see in the next sections.

In the study of semiclassical spin tunnelling, the relevant
classical trajectories are instantons with the remarkable property
that $\bz(0)=z_i^\ast$ and $z(T)=z_f$ \cite{instanton}. This
happens because the initial and final points of the instanton
minimize the average value of the Hamiltonian, and in that case
the calculation is greatly simplified, leading to analytical
results. However, as noted in \cite{jumps} a more realistic
description of the system requires the addition of higher order
terms to the Hamiltonian that destroy this simple property and
make it necessary to consider the more generic trajectories we
have discussed (which have been called ``boundary jump instantons"
in \cite{jumps}).

Note that the action (\ref{action}), with the necessary boundary
term, leads to the following Hamilton-Jacobi relations
\begin{align}\label{HJ} \frac{i}{\hbar}\frac{\partial S}{\partial
\bz_f}=\frac{2jz(t)}{1+\bz_f z(t)},& \quad
\frac{i}{\hbar}\frac{\partial S}{\partial
z_i}=\frac{2j\bz(0)}{1+\bz(0)z_i}, \\ \frac{\partial S}{\partial
t}&=-H.\end{align} Note also that if the Hamiltonian is $O(j)$,
then $S$ is $O(j)$, but the SK correction is $O(1)$, and therefore
can be considered small in the semiclassical limit.

The prefactor is related to the tangent matrix, or stability
matrix, as follows. Small variations in the boundary points
$\delta z_i$ and $\delta \bz_f $ induce variations $\delta z(T)$
and $\delta \bz(0)$. Taking derivatives of the Hamilton-Jacobi
relations (\ref{HJ}) it is possible to write \be
\label{As}\left(\begin{array}{c}
  (1+\bz_f z(T))^{-2}\delta z(T) \\
  (1+z_i \bz(0))^{-2}\delta \bz(0)\end{array} \right)=\begin{pmatrix}
    A_{zz} & A_{z\bz} \\
    A_{\bz z} & A_{\bz\bz} \end{pmatrix}\left(\begin{array}{c}
  \delta z_i \\
  \delta \bz_f\end{array} \right),
\ee where all matrix elements can be written in terms of second
derivatives of the action (see \cite{jpa34mb2001} for an analogous
calculation with canonical coherent states). On the other hand,
the tangent matrix of a given trajectory $(z(t),\bz(t))$ is
defined as the linear application that takes a small initial
displacement to a final displacement, \be \left(\begin{array}{c}
  \delta z(T) \\
  \delta \bz_f\end{array} \right)=\begin{pmatrix}
    M_{zz} & M_{z\bz} \\
    M_{\bz z} & M_{\bz\bz} \end{pmatrix}\left(\begin{array}{c}
  \delta z_i \\
  \delta \bz(0)\end{array} \right).\ee Manipulating
  equation (\ref{As}) one can show that the element $M_{\bz\bz}$ is
  related to the prefactor in (\ref{propag}) according to
  \be\label{caus} \frac{i}{\hbar}\frac{\partial^2 S}{\partial z_i\partial
\bz_f}=\frac{2j}{(1+z_i\bz(0))^2}\frac{1}{M_{\bz\bz}}. \ee Since
the tangent matrix may be numerically integrated together with the
coordinates, this expression is very convenient in practice. The
phase $\varphi_p$ may then be followed dynamically, imposing that
$K=1$ ($\varphi_p=0$) for $T=0$.

In practice, one must find all values of $\bz(0)$ for which
$\bz(T)=z_f^{\ast}$. Notice that these initial values are implicit
functions of the time $T$, and as such they trace out curves, or
``branches", in the $\bz(0)$-plane. There will in general exist
more than one branch for each value of $T$, and one must add all
contributions coherently. Note, however, that the contribution of
a branch may display an erroneous increase and need to be removed
for a certain region of the parameter $T$, because of a phenomenon
that is common in asymptotic expansions known as Stokes'
phenomenon \cite{stokes,dingle,berry,shudo}. A discussion of this
problem in the context of the one-dimensional semiclassical
propagator may be found in \cite{heller}. Stokes' phenomenon will
not be relevant to the present work.

From equation (\ref{caus}) it is clear that the semiclassical
approximation fails in the vicinity of points for which
$M_{\bz\bz}(T)=0$, which are called phase-space caustics
\cite{pre69adr2004,heller,caustics}. Semiclassical uniform
approximations that are valid near caustics have been obtained for
the canonical coherent state propagator using a conjugate of the
Bargmann representation in \cite{uniform}, and a similar
calculation is in principle possible for the spin propagator, but
here we shall not be concerned with the effect of caustics.

\section{First example, $\hat{H}= \hbar^2\nu J^2_z$}

We consider a simple Hamiltonian, \be H=\hbar^2\nu J^2_z,\ee (the
constant $\nu$ has the appropriate units) and we shall be
interested only in the diagonal propagator \be K(z_i,T)=\langle
z_i| e^{-iHT/\hbar}|z_i\rangle,\ee whose squared modulus, after
proper normalization, corresponds to the return probability as a
function of time. This is given by \be K(z_i,T)=\sum_{m=-j}^j
|\langle m|z_i\rangle|^2e^{-im^2\hbar \nu T},\ee which for integer
$j$ is periodic with period $T_r=2\pi/\hbar \nu$. In the
semiclassical limit the term that contributes the most to this sum
(see (\ref{states})) is $m_0=j(|z_i|^2-1)/(|z_i|^2+1)$. If we
linearize the exponent in the vicinity of this term we have
\be\label{linear} K(z_i,\tau)\approx|\langle
m_0|z_i\rangle|^2e^{-im_0^2\tau}\sum_{n\approx
m_0}e^{-2im_0n\tau}, \ee where we have introduced a scaled time
\be \tau=\hbar \nu T.\ee Notice that expression (\ref{linear}) has
a different time scale, $\tau_c=\pi/m_0$. The quantities $T_r$ and
$T_c=\tau_c/\hbar \nu$ are usually called revival time and
classical time \cite{revivals}.

Let us turn to the semiclassical approximation, which has been
analyzed in some previous works \cite{koche2,vieira,jmp41ms2000}.
The classical Hamiltonian is \be H=\hbar^2\nu
j(j-\frac{1}{2})\left(\frac{z\bz-1}{z\bz+1}\right)^2+\frac{\hbar^2\nu
j}{2}.\ee The classical equations of motion,
$\dot{z}=-i\hbar\nu\omega z$ and $\dot{\bz}=i\hbar\nu\omega \bz$
(here a dot denotes derivative with respect to $t$), together with
the boundary conditions $z(0)=z_i$ and $\bz(T)=z_i^\ast$, have the
simple solutions \be z(t)=e^{-i\hbar\nu\w t}z(0), \quad
\bz(t)=e^{i\hbar\nu\w t}\bz(0).\ee Notice that \be \label{omega}
\w=\frac{2\mu}{j}\left(\frac{z\bz-1}{z\bz+1}\right)\ee is a
constant of the motion, where $\mu=j(j-1/2)$. Calculating it at
the initial and final points we have the consistency condition
\be\label{consis} \w=\frac{2\mu}{j}\left(\frac{e^{-i\w
\tau}|z_i|^2-1}{e^{-i\w \tau}|z_i|^2+1}\right),\ee which in
general has an infinite number of solutions. We come back to that
later. Notice also that for $z(0)=z_i$ and $\bz(0)=z_i^\ast$ the
motion is periodic with period $2\pi/\hbar\nu\w$, which in the
semiclassical limit becomes $T_c$.

The action can be easily found to be \be
\frac{S}{\hbar}=-2ij\ln(1+e^{-i\w
\tau}|z_i|^2)+\tau\left[j\w+\frac{j^2\w^2}{4\mu}-\frac{j}{2}\right],\ee
and the Solari-Kochetov term is also available, \be
A=\hbar^2\nu\left[\frac{j\w+\mu}{2j}-\frac{j\w^2}{8\mu}\right].\ee
To find the prefactor, we consider small variations $\delta z_i$
and $\delta \bz(0)$. The final value of $\bz$ will be changed
according to
\begin{align}
\bz(\tau)+\delta\bz(\tau)&=e^{i(\w+\delta\w)\tau}(\bz(0)+\delta
\bz(0))\nonumber\\&\approx \bz(\tau)+M_{\bz\bz}\delta
\bz(0),\end{align} which leads to \be M_{\bz\bz}=e^{i\w
\tau}\left[1+\frac{4\mu}{j}\frac{i\tau
z_i\bz(0)}{(1+z_i\bz(0))^2}\right].\ee There are two different
phase space caustics, that can be found by equating
$M_{\bz\bz}=0$. They are located, as functions of the scaled time,
along the curves
$\bz_\pm(0)=-z_i^{-1}\left(1+i\tau\mp\sqrt{2i\tau-\tau^2}\right)$.
We shall not be concerned here with the effect of caustics.

The problem now is to find the values of $\bz(0)$ for which
$\bz(T)=z_i^\ast$. These points form curves in the plane $\bz(0)$,
that we call branches and denote $\bz_{\tau}(0)$, since they are
parametrized by the time. We note again that even for a fixed time
there may exist many branches. The method used for finding them
was the following: for a fixed value of time, $\tau_0$, a regular
grid is placed on the $\bz(0)$ plane, and each point is taken as
an initial condition. Those points for which $\bz(\tau_0)$ is
close to $z_i^\ast$ are then used as initial guesses for a
root-finding procedure. For $\tau=\tau_0\pm\delta \tau$ the
previously obtained solutions serve as initial guesses, and no
grid is used. This way the branches may be obtained rapidly and
exhaustively. Of course the point $\bz(0)=z_i^\ast$ is a solution
for $\tau=0$, and thus it must be contained in one of the
branches.

\begin{figure}[t]
\includegraphics[scale=0.29,angle=-90]{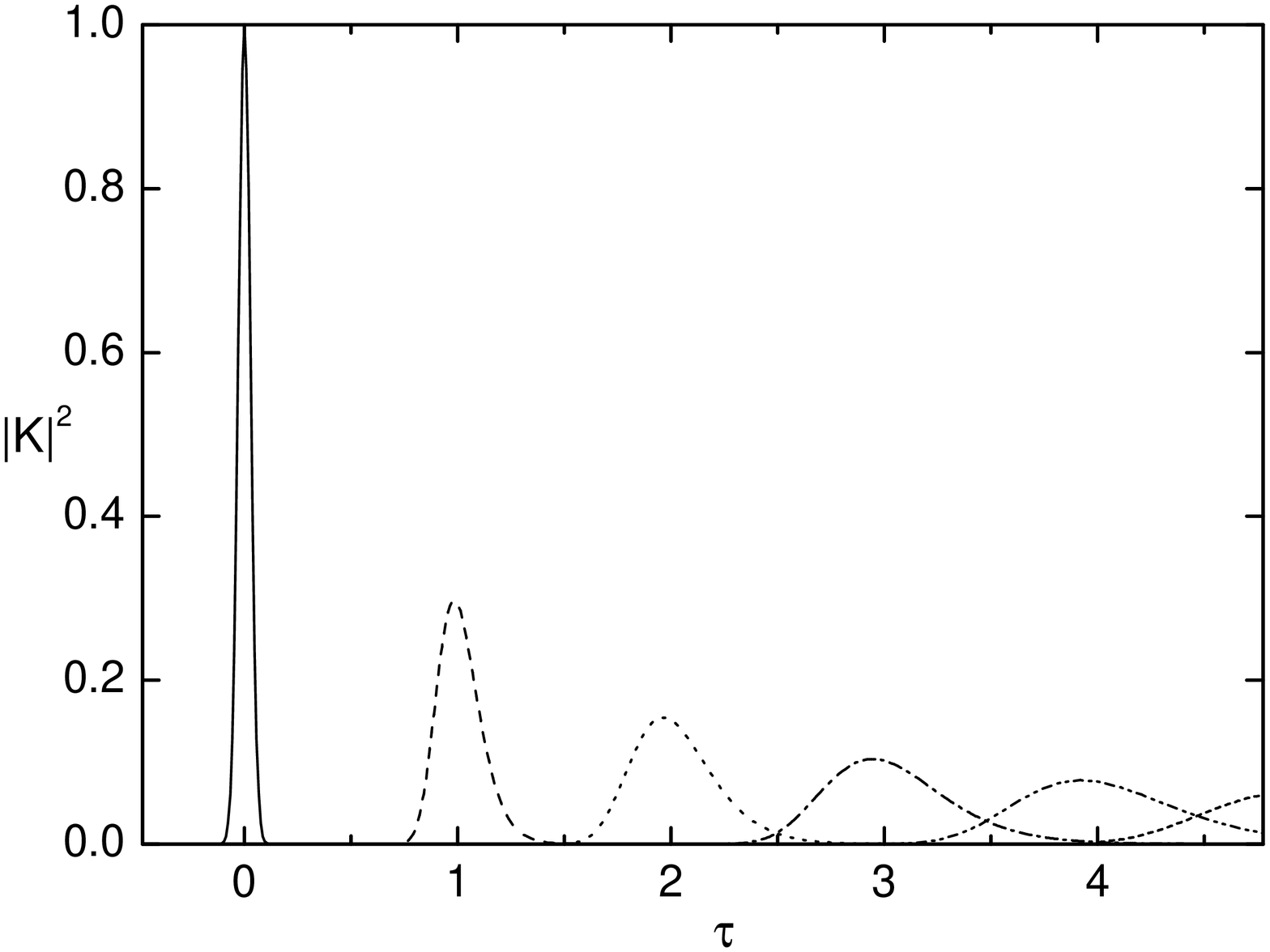}\\
\includegraphics[scale=0.29,angle=-90]{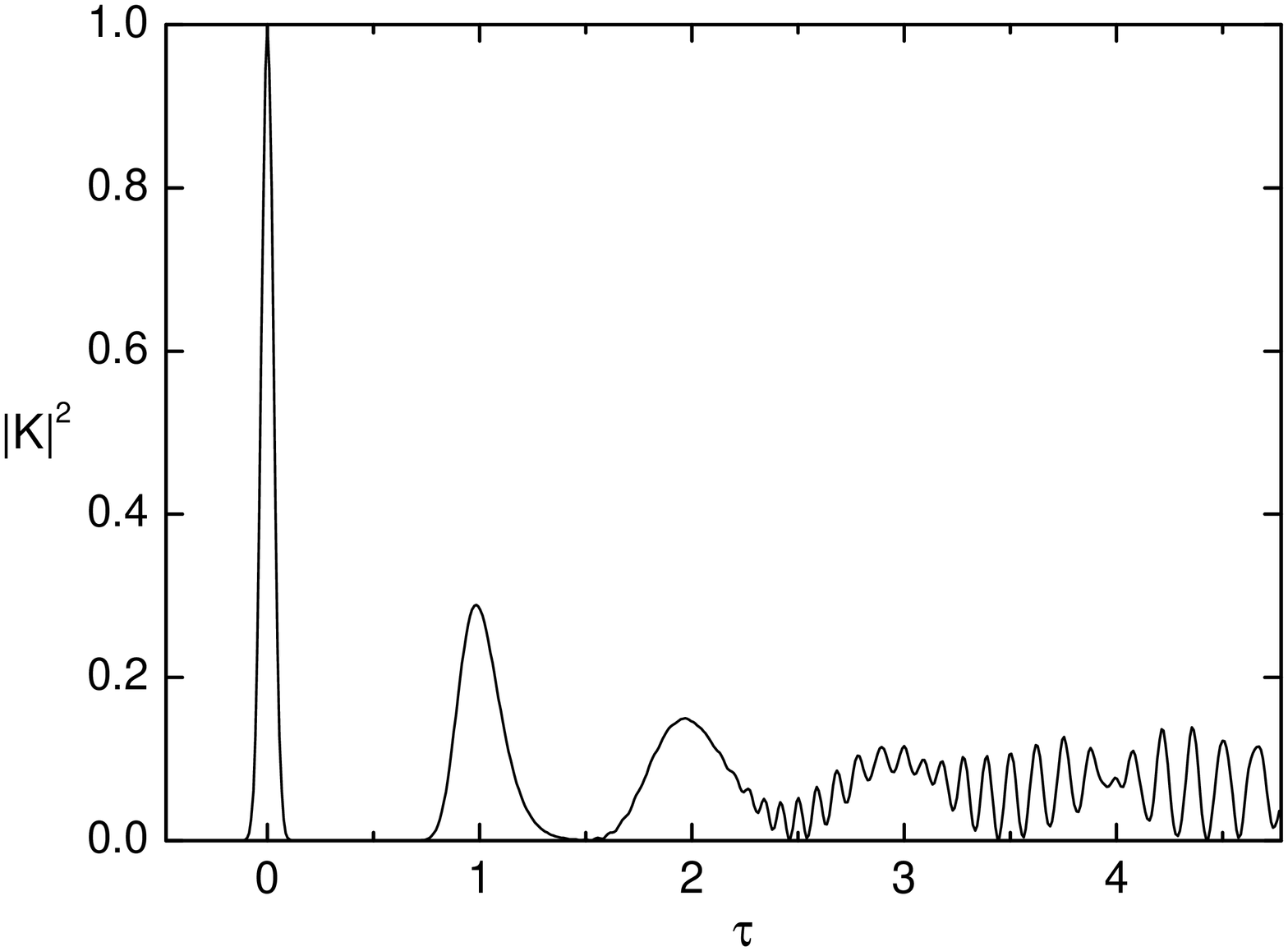}
\caption{The normalized propagator as a function of scaled time
$\tau=\hbar\nu T$ (in units of the classical period $\tau_c$) for
$z_i=0.5$ and $j=50$. In (a) we see the separate contributions of
$6$ branches (each peak is due to a different branch). When they
are added coherently we get (b), which is indistinguishable from
the exact result, including the fast oscillations.}
\end{figure}

\begin{figure}[t]
\includegraphics[scale=0.29,angle=-90]{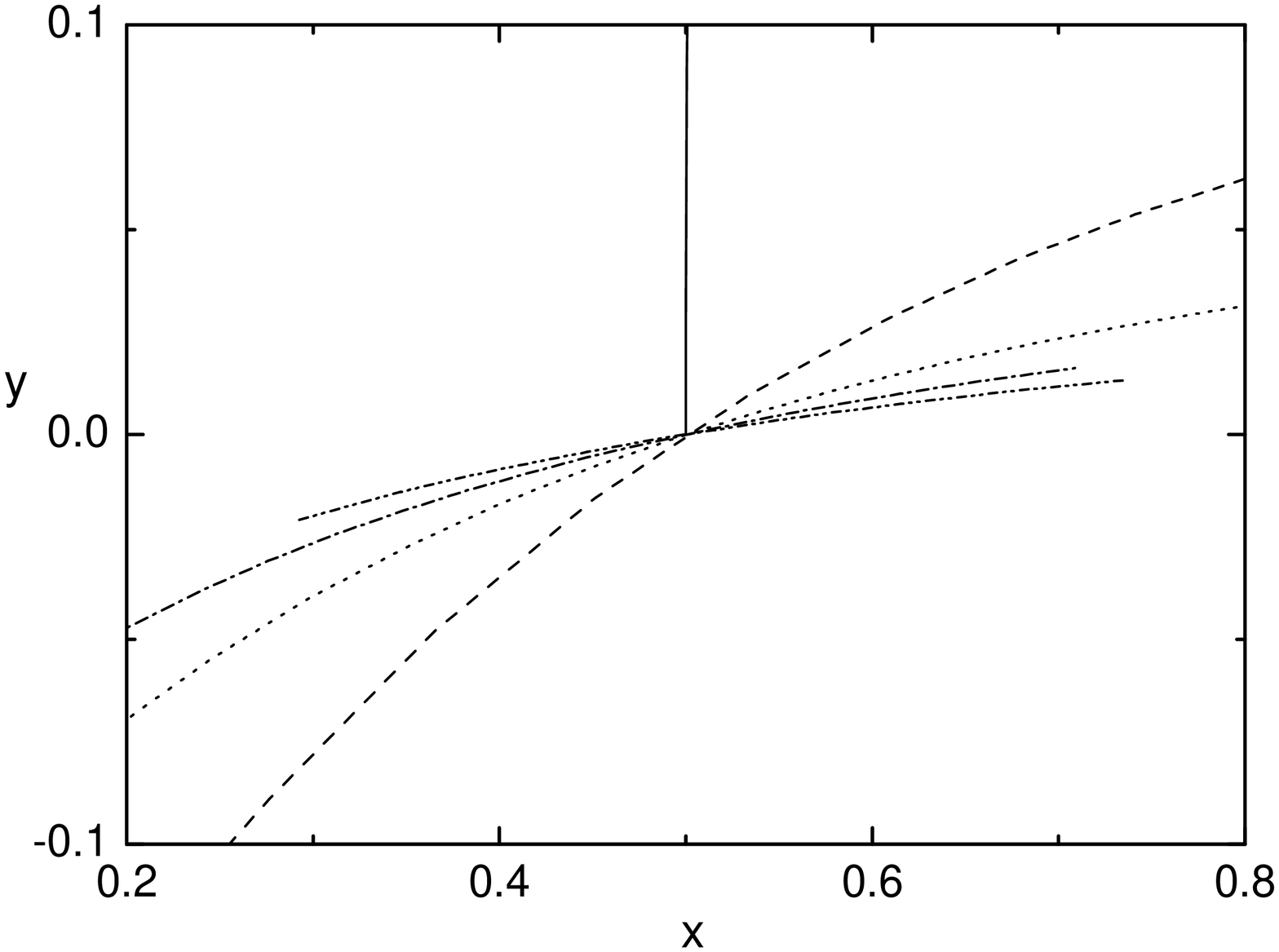}
\caption{The different branches in the $\bz(0)=x+iy$ plane
contributing to the semiclassical return probability, for the same
parameters as in the previous figure (quantities are
dimensionless). These curves as parametrized by the time, and each
one of them passes through the point $(0.5,0)$ at a different
instant. Only the vicinity of this point is shown, for the sake of
visibility. }
\end{figure}

As an example, let us consider $z_i=0.5$ and a very semiclassical
regime, $j=50$ (which gives $\hbar=0.02$). For the time interval
of five classical periods we have found $6$ different branches.
The squared modulus of each of their normalized contributions is
shown in Fig. 1a, with different line styles. We see that each
peak in the return probability is due to a different branch. The
appearance of the classical time scale $\tau_c\approx 0.1$, which
comes from $m_0=-30$, is very clear. For $\tau\approx 0$ the
branch depicted in solid line is in the vicinity of the point
$z_i^\ast$, and as $\tau$ approaches the integer multiples of
$\tau_c$ each one of the other branches approaches this point,
generating a peak in the return probability. The behavior of the
different branches can be seen in Fig. 2. After the third peak the
contributions start to overlap, and therefore we must add them
coherently to get interference effects. When this is done, the
result is as shown in Fig. 1b. The exact calculation is easy to
perform, and it is indistinguishable from the semiclassical one at
this scale. Even the fast oscillations for $\tau>2.5\tau_c$ are
reproduced.

When the value of $z_i$ corresponds to a point along the equator
of the Bloch sphere, i.e. when $|z_i|=1$, the calculation may be
simplified. In that case there exists a very particular classical
trajectory: the one for which $z(t)=z_i$ and $\bz(t)=z_i^\ast$. In
this case it easy to see that $\w=0$ and there is no movement. Let
us call this the static trajectory. Each different value of
$\bz_{\tau}(0)$ determines a certain $\w$, and the final
expression for the normalized propagator is
\begin{align}\label{equator}& K_{\rm sc}(z_i,\tau) =\sum_\w\left[1-\frac{i\tau}{4\mu
j}(\w^2j^2-4\mu^2)\right]^{-\frac{1}{2}}\nonumber\\& \times
\cos^{2j}\left(\frac{\w \tau}{2}\right)\exp \left\{\frac{i\tau
j}{2}\left(
\frac{\w^2}{2j}+\frac{\mu}{j^2}-1\right)\right\},\end{align} where
we have used (\ref{omega}) and the fact that $z\bz$ is a constant
in time. It is easy to see that each term in this sum (as well as
equation (\ref{consis})) is even in $\w$, and therefore that
trajectories must come in pairs. The net effect is that we may
search only for $\w$'s with a positive real part, and once a
certain trajectory is found its contribution must be doubled,
except for the static one, which gives \be |K_{\rm
st}(z_i,\tau)|^2=\left(1+\tau^2(j-\frac{1}{2})^2\right)^{-\frac{1}{2}}.\ee

\begin{figure}[t]
\includegraphics[scale=0.29,angle=-90]{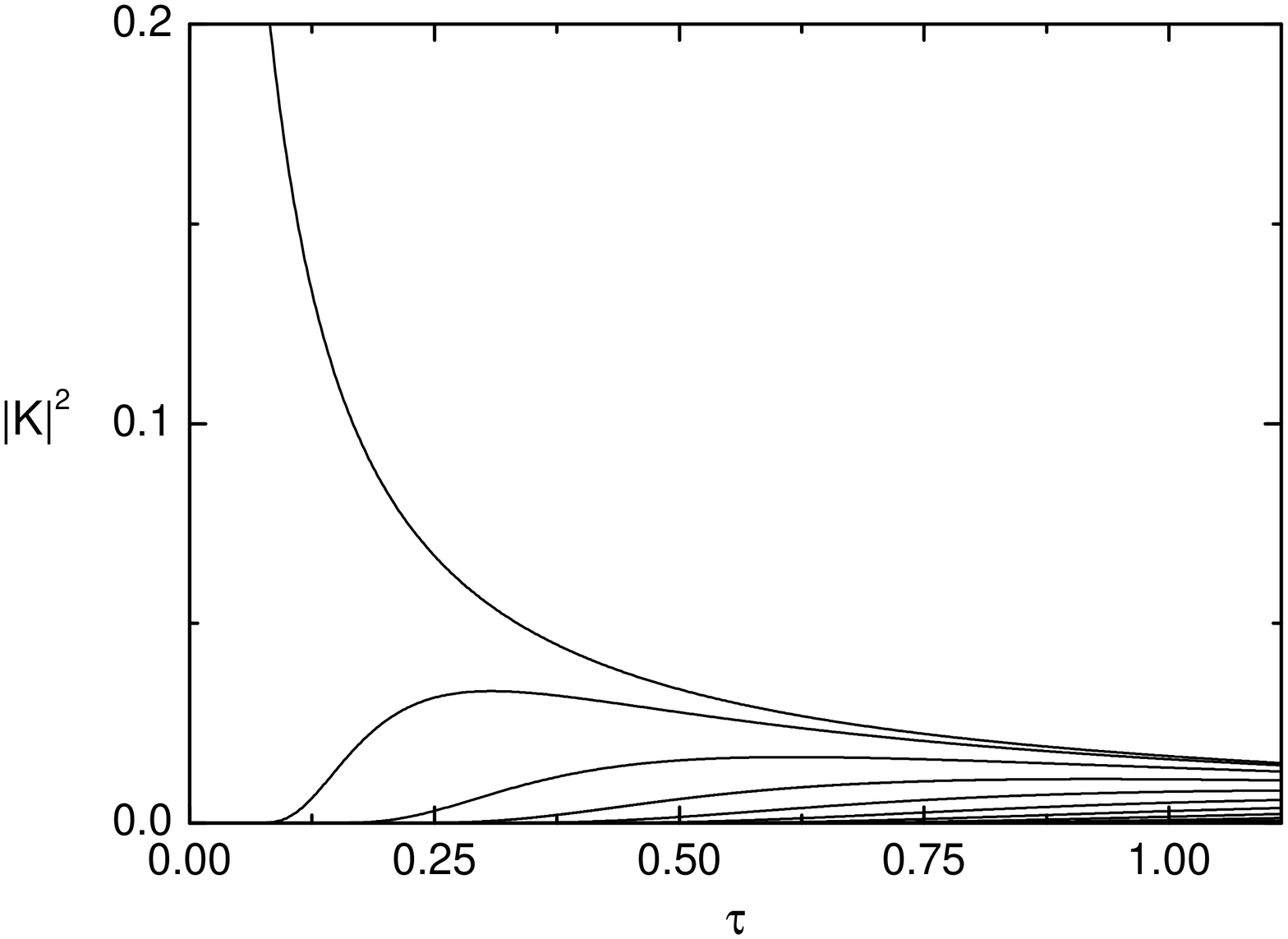}\\
\includegraphics[scale=0.29,angle=-90]{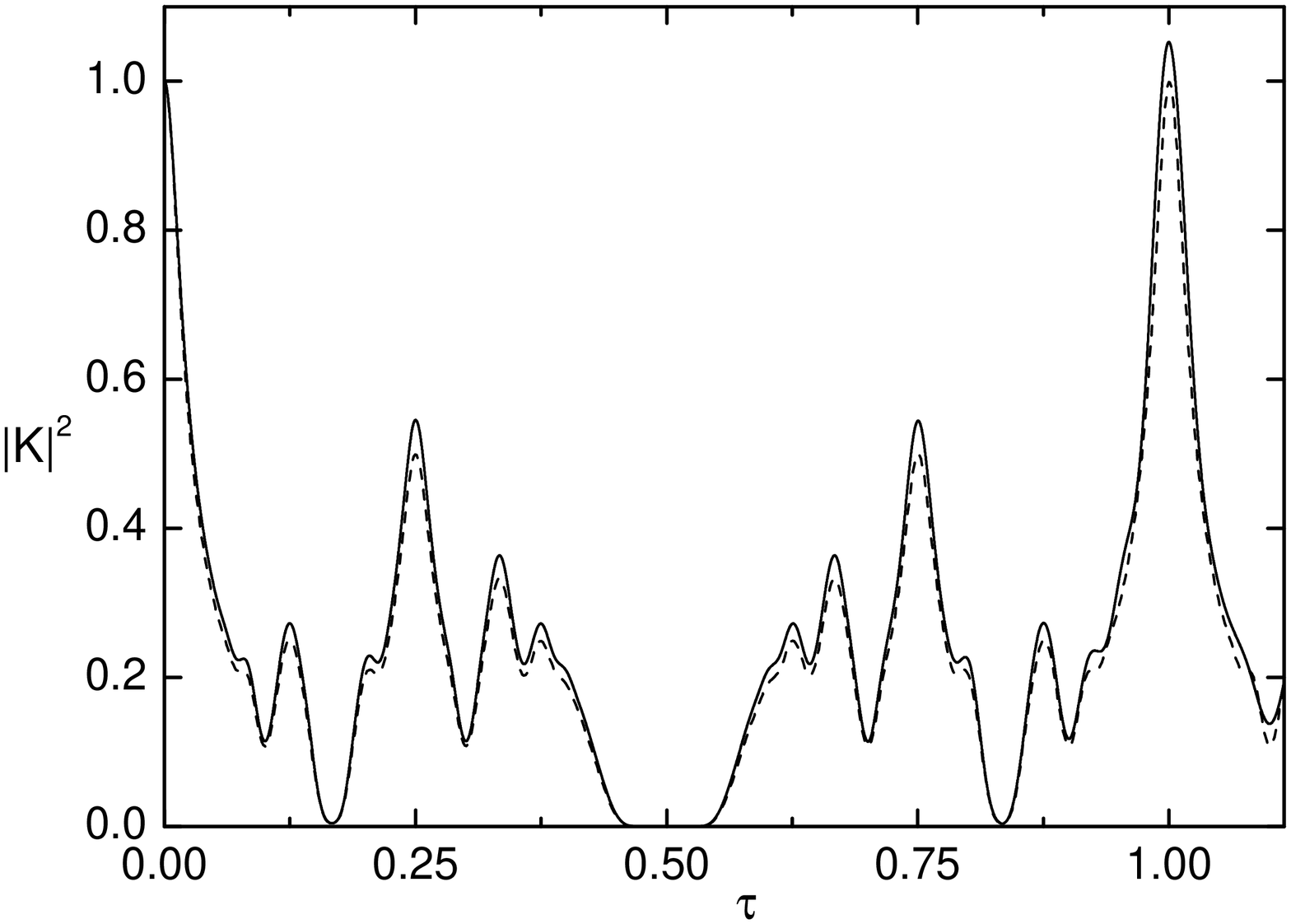}
\caption{The normalized propagator as a function of scaled time
(in units of $\tau_r$) for $z_i=1$ and $j=10$. In (a) we see the
separate contributions of $11$ branches (all of them except one
must be counted twice, see text). When they are added coherently
we get the solid line in (b). The dashed line is the exact result.
}
\end{figure}

For short times we may approximate \be |K_{\rm
st}(z_i,\tau)|^2\approx 1-\frac{\tau^2}{2}(j-\frac{1}{2})^2,\ee
and comparing this with the exact short-time return probability
\begin{align} |K(z_i,\tau)|^2&\approx 1-\left(\langle
H^2\rangle-\langle H\rangle^2\right)
T^2/\hbar^2\nonumber\\&=1-\frac{\tau^2}{2}j(j-\frac{1}{2}),\end{align}
(where $\langle \cdot\rangle=\langle z_i|\cdot|z_i\rangle$) we see
that in the semiclassical limit the short-time regime is well
described by the static trajectory alone.

As a concrete example, we have considered $z_i=1$ and $j=10$, so
that higher order quantum effects may become more visible. In this
case $11$ branches were found that contributed to the final
result. Their separate contributions to the normalized propagator
can be seen in Fig. 3a as functions of time (in units of the
revival time $\tau_r$). Because of the even parity in $\w$, each
one of them was considered twice, except for the static one. When
we add them coherently, we get the solid line in Fig. 3b, which is
to be compared with the exact result, the dashed line. We see that
the later is reproduced with an extraordinary accuracy, including
the quantum revival.

\begin{figure}[t]
\includegraphics[scale=0.29,angle=-90]{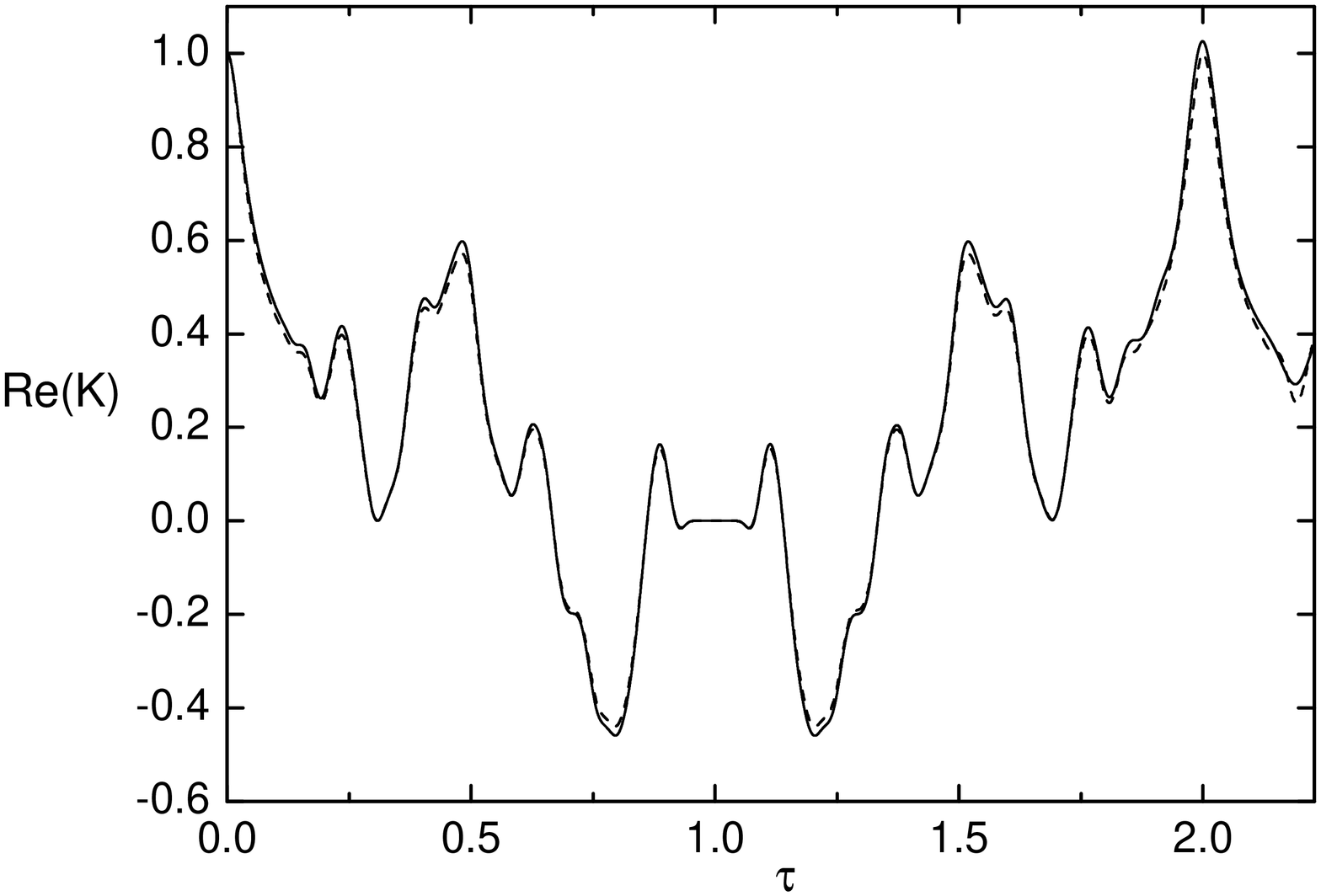}
\caption{The semiclassical (solid) and the exact (dashed)
calculations of the real part of the normalized propagator, for
$z_i=1$ and $j=10$. The imaginary part is also reproduced with
great accuracy. Therefore the total phase, dynamical plus
geometrical terms, can be recovered from the semiclassical
expressions.}
\end{figure}

We do not show the branches in the $\bz(0)$ plane, but their
behavior is quite simple. The static branch lyes in the point
$\bz(0)=1$ for all times. For short times all other branches are
at the vicinity of $-1$, but their contribution to the propagator
is negligible. As time passes, they move away from this point and
towards $1$. The most important thing to note is that all branches
have appreciable contributions at the period $\tau=2\pi$, so the
phase of the propagator is determined by their coherent
superposition and has no simple expression.

Concerning the phase in (\ref{equator}), notice that it can be
written as $i\tau(\w^2-1)/4$ because $\mu/j^2-1$ is actually equal
to $-1/2j$. Since $\tau$ is of order $\hbar$ the second term
vanishes in the semiclassical limit, but the first one remains
because $\w^2\sim j^2$. However, we have observed numerically that
the factor $e^{-i\tau/4}$, which is common to all branches (and
thus has no relevance to the modulus), destroys the correspondence
with the exact result. Once this term is removed, which is
equivalent to the semiclassically acceptable approximation
\be\label{approx}\mu=j(j-1/2)\approx j^2,\ee the agreement is
excellent, as we can see from Fig. 4, where the real part of the
normalized propagator is shown (the imaginary part is also very
accurate). This approximation must not be done in the prefactor,
where we have used $\mu=j(j-1/2)$.

\begin{figure}[t]
\includegraphics[scale=0.29,angle=-90]{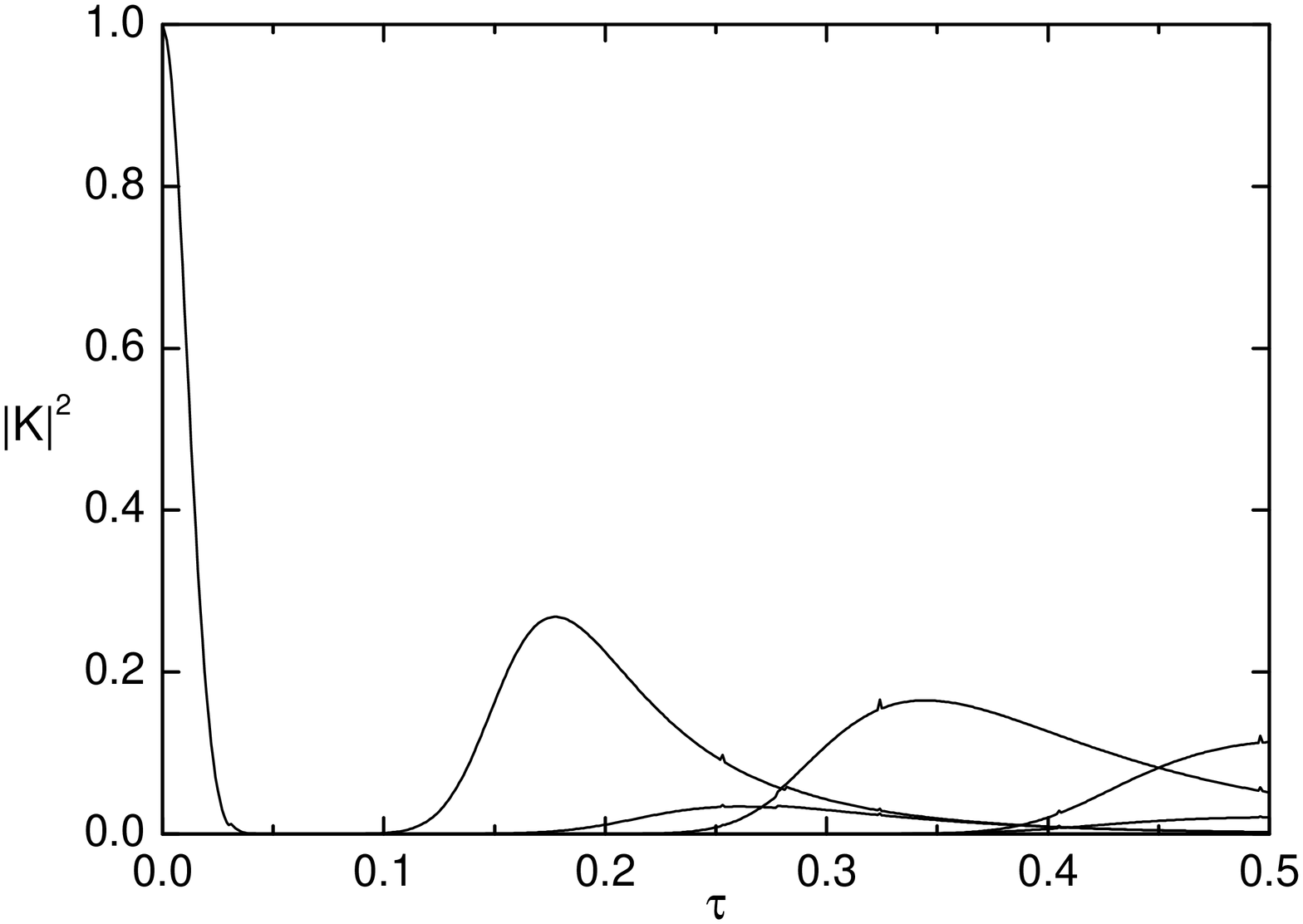}\\
\includegraphics[scale=0.29,angle=-90]{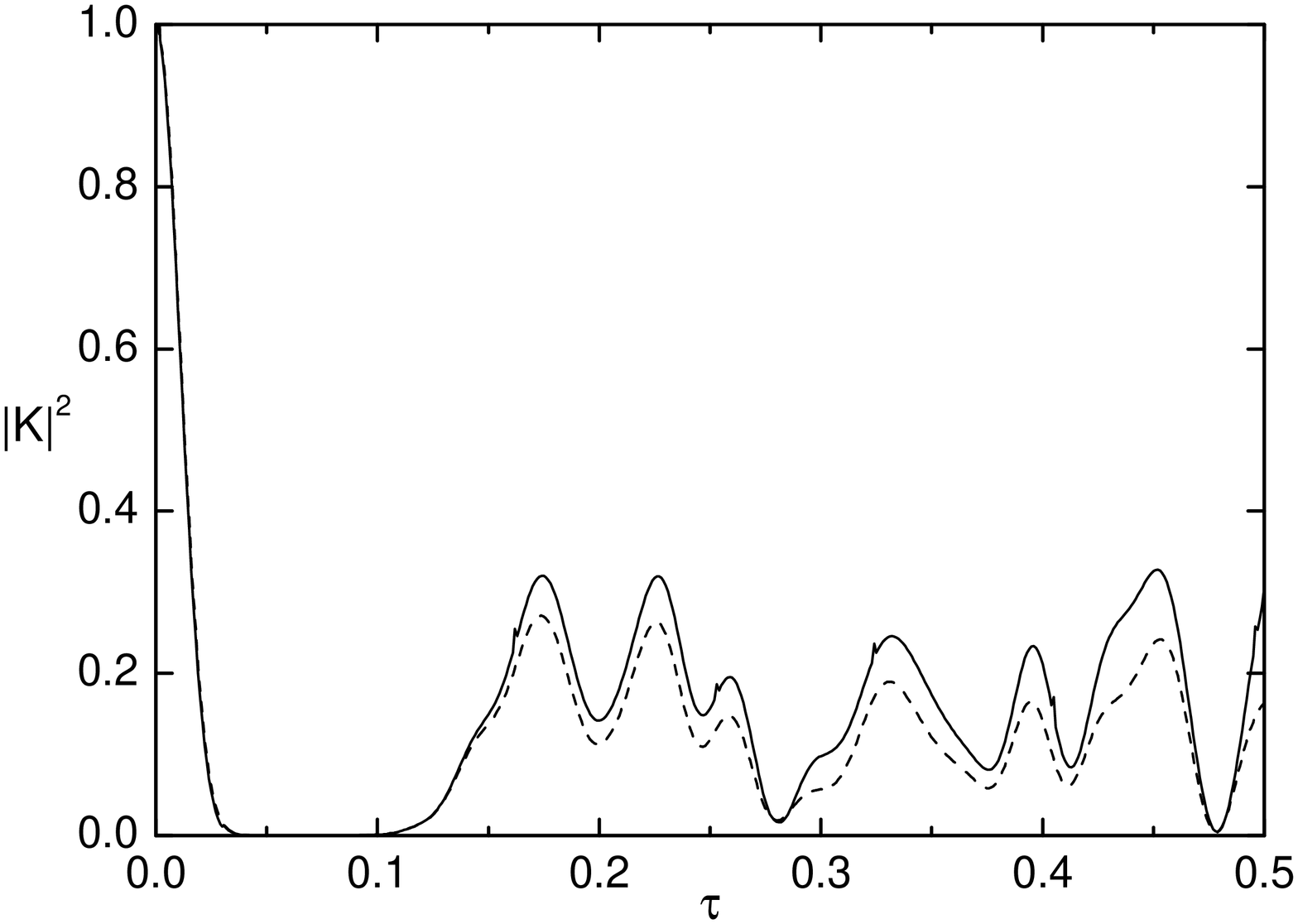}
\caption{The normalized propagator as a function of scaled time
(in arbitrary units) for $z_i=1+i$ and $j=10$ in the second
example. In (a) we see the separate contributions of $6$ branches.
When they are added coherently we get the solid line in (b). The
dashed line is the exact result. }
\end{figure}

We therefore conclude that expression (\ref{propag}) is able to
reproduce both modulus and total phase, dynamical and geometric
terms combined, of the quantum propagator with great accuracy in
the semiclassical limit. It involves a coherent sum over classical
trajectories, and thus it is not possible to extract the geometric
phase from it in a closed form. As noted in the previous section,
the geometric phase will be given by (\ref{geo}) only in the
simplest cases, when a single classical trajectory is involved.
Even then it will in general be different from the simple formula
$\int_0^Tijdt(\bz\dot{z}-z\dot{\bz})/(1+z\bz)$.

Stone \emph{et al} have analyzed the case $\hat{H}=J^2_z$ in
detail \cite{jmp41ms2000}. They have shown the formal equivalence
of the semiclassical and exact (non-normalized) propagators up to
order $j^0$, and also that for $|z_i|=1$ the result for a massive
particle constrained to move on a ring is recovered as
$j\to\infty$. However, in \cite{jmp41ms2000} no numerical
calculation has been done, and the fact that replacing $\mu$ by
$j^2$ in the phase improves the result was not noticed.

\section{Second example, $\hat{H}=\hbar^2\nu[J^2_z+\pi J^2_x]$}

As a second example, we choose the less symmetrical Hamiltonian
$\hat{H}=\hbar^2\nu[J^2_z+\kappa J^2_x]$, which has been
considered in \cite{prl60emc1988,vieira}. We have chosen the
irrational value $\kappa=\pi$, in order to avoid too much
simplicity. The exact calculation of the return probability can be
done by direct diagonalization of the Hamiltonian in the $J_z$
basis. The initial state was taken as a complex number out of the
equator, $z_i=1+i$. For very large values of $j$ the return
probability is very similar to what we see in Fig. 1b, but for
more moderate values it is much less regular.

\begin{figure}[t]
\includegraphics[scale=0.29,angle=-90]{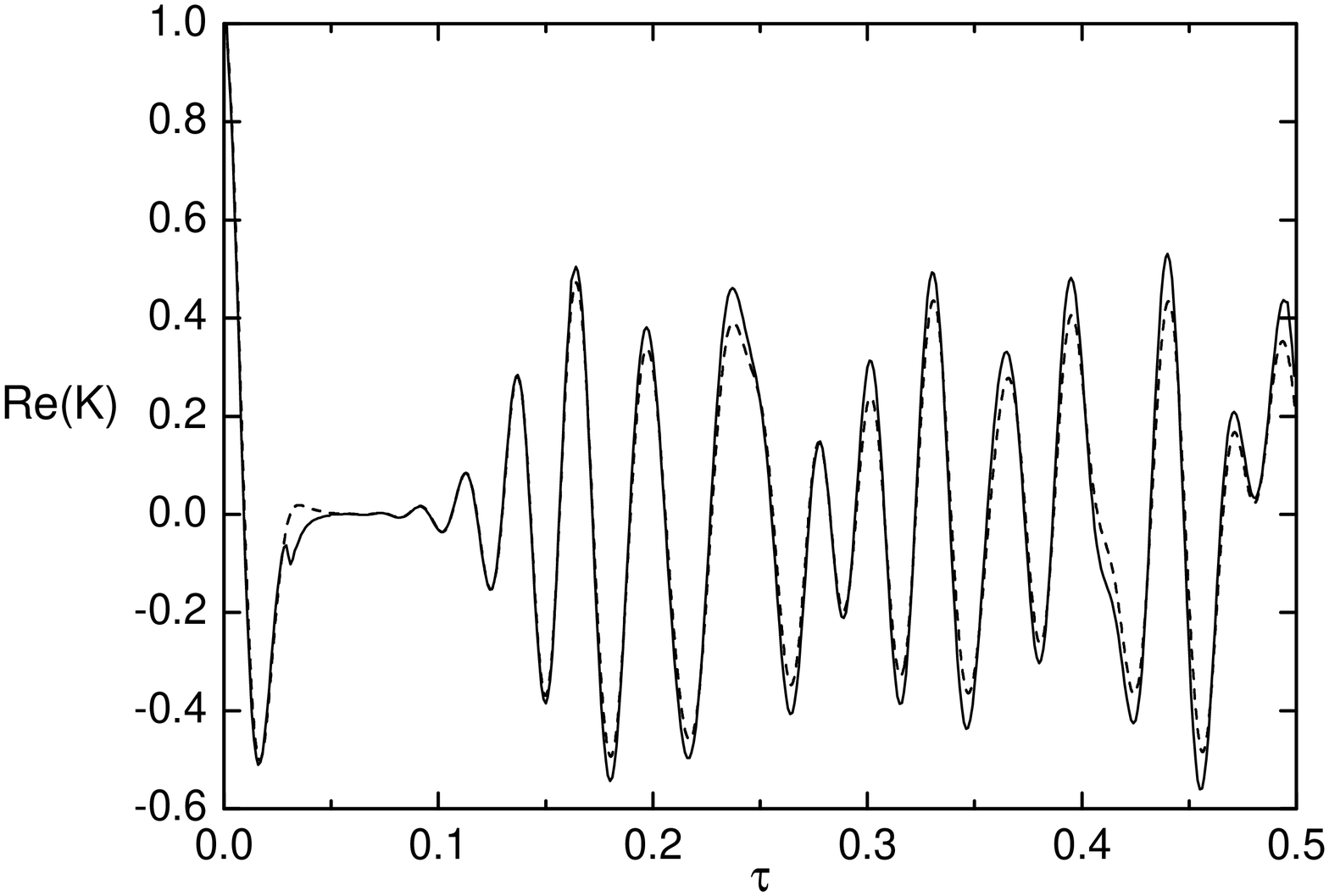}
\caption{The semiclassical (solid) and the exact (dashed)
calculations of the real part of the normalized propagator, in the
same situation as the previous figure. Once again the phase is
accurately reproduced.}
\end{figure}

In Fig.5a we see the separate contribution of $6$ branches, each
of which has only one peak. This time the classical trajectories,
the tangent matrix, the action and the SK correction all had to be
determined through a numerical integration, using a Runge-Kutta
method. The coherent superposition of the individual contributions
is the solid line in Fig.5b, and the dashed line is the exact
result. Even though the dynamics does not show any periodicity or
regularity in the considered time interval, the semiclassical
approximation works very well. Notice that we have used the same
scaled time as in the previous example, $\tau=\hbar\nu T$.

The phase of the propagator is again recovered with great
precision, as we see in Fig.6, where the real part of both the
semiclassical and the exact calculations is shown (again the
imaginary part is reproduced with the same accuracy). This time no
approximation of the kind (\ref{approx}) was made to the phase,
which was obtained numerically.

\section{Conclusions}

A numerical evaluation of the semiclassical spin coherent state
propagator has been presented for the first time. We have
considered two different systems as examples, the quantum rotor
$\hat{H}=\hbar^2\nu J^2_z$ and the less symmetrical Hamiltonian
$\hat{H}=\hbar^2\nu[J^2_z+\pi J^2_x]$. In the first case the exact
calculation of the return probability is very simple, and is
reproduced by the semiclassical one with excellent precision for
$j=50$. For the smaller value $j=10$ the accuracy of the
approximation is still remarkable.

We have made the numerical observation that a factor $-i\hbar\nu
T/4$ must be removed from the semiclassical phase in order to
reproduce the exact quantum phase for the rotor (the modulus is
not affected). We have no rigorous explanation for this fact,
except that this term vanishes in the semiclassical limit and that
its removal corresponds to the replacement $j(j-1/2)\to j^2$.
Since phases are defined modulo $2\pi$, we believe this limit must
be considered with great care. Since no such adjustment had to be
made for the second Hamiltonian studied, we conclude that
geometric phases are adequately reproduced by semiclassical
calculations. However, they generally do not have a simple closed
formula, but result from the interference of many classical paths.

The most important difficulty in the implementation of the
semiclassical spin propagator in practice is finding the relevant
classical trajectories, i.e. those values of $\bz(0)$ for which
the boundary condition at time $T$ is satisfied,
$\bz(T)=z_f^\ast$. We have done this by first propagating for a
time $T$ a grid in the $\bz(0)$ plane, thus finding initial rough
estimates for the relevant points. These were then used to feed a
root-finding procedure. The whole process is rather artisanal and
hard to automatize, making it almost impossible to tackle problems
in which a large numbers of trajectories is necessary. The same
kind of difficulty is found in the case of canonical coherent
states.

Another problem, which we have not considered, is the existence of
caustics, initial conditions for which the prefactor diverges at
time $T$. We have seen that for $\hat{H}=\hbar^2\nu J^2_z$ there
are two such points, but they did not have any influence in the
cases we have analyzed. For systems with higher nonlinearities it
is reasonable to expect a larger number of caustics, which could
also hinder the practical application of the semiclassical
approximation. In \cite{uniform} the authors have developed an
approach to treat the vicinity of caustics in the canonical
coherent states propagator, that makes use of a certain dual of
the Bargmann representation. This could in principle be
generalized to the spin case.

The main virtue of the semiclassical propagator is its weak
dependence on the dimension of the representation, $2j+1$. If this
is too large a direct diagonalization of the Hamiltonian may
become impracticable, while the difficulty with the semiclassical
method is basically the same. Therefore an excellent approximation
to the exact result, at least for short times, would be easily
available.

\section*{Acknowledgments} Financial support from
FAPESP (Funda\c{c}\~ao de Amparo \`{a} Pesquisa do Estado de S\~ao
Paulo) is gratefully acknowledged. I wish to thank M.A.M. de
Aguiar, A.D. Ribeiro and F. Parisio for important discussions.

\end{document}